\PassOptionsToPackage{table}{xcolor}
\documentclass[acmtog, nonacm]{acmart}

\makeatletter                   
\def\mdseries@tt{m}             
\makeatother                    

\usepackage{booktabs} 
\usepackage{amsmath, amssymb, bm}
\usepackage{amsfonts}
\usepackage{hyperref}
\usepackage{color}
\usepackage{mathtools}
\usepackage[inline]{enumitem}
\usepackage{tabularx}

\setcitestyle{square}

\definecolor{myorchid}{RGB}{10,10,200}
\definecolor{darkgreen}{RGB}{10,150,10}
\definecolor{darkred}{RGB}{127,10,20}
\definecolor{YellowOrange}{RGB}{255,127,39}

\newcommand{\comm}[3]{}

\newcommand{\ud}[1]{#1} 

\newcommand{\BR}{\mathbb{R}}
\newcommand{\br}{\mathbf{r}}

\newcommand{\bv}{\mathbf{v}}
\newcommand{\bx}{\mathbf{x}}
\newcommand{\by}{\mathbf{y}}
\newcommand{\be}{\mathbf{e}}

\newcommand{\cN}{\mathcal{N}}
\newcommand{\cE}{\mathcal{E}}
\newcommand{\cM}{\mathcal{M}}
\newcommand{\cV}{\mathcal{V}}
\newcommand{\cT}{\mathcal{T}}

\newcommand{\BZ}{\mathbb{Z}}
\newcommand{\bG}{\mathbf{G}}
\newcommand{\tphi}{\widetilde{\phi}}

\DeclareMathOperator{\argmin}{argmin}

\DeclareMathOperator{\expm}{expm}

\makeatletter
\newcommand{\doublewidetilde}[1]{{%
  \mathpalette\double@widetilde{#1}%
}}
\newcommand{\double@widetilde}[2]{%
  \sbox\z@{$\m@th#1\widetilde{#2}$}%
  \ht\z@=.9\ht\z@
  \widetilde{\box\z@}%
}
\makeatother

\usepackage[ruled]{algorithm2e} 

\SetAlFnt{\small}
\SetAlCapFnt{\small}
\SetAlCapNameFnt{\small}
\SetAlCapHSkip{0pt}
\IncMargin{-\parindent}

\usepackage{subcaption}	
\usepackage{wrapfig}

\setcopyright{none}

\acmDOI{0000001.0000001_2}

\begin{document}
\title{\ud{Designing Volumetric Truss Structures}}

\author{Rahul Arora}
\orcid{0000-0001-7281-8117} 
\affiliation{%
 \institution{University of Toronto}
 \streetaddress{40 St.~George Street}
 \city{Toronto}
 \state{ON}
 \postcode{M5S~2E4}
}
\author{Alec Jacobson}
\affiliation{%
 \institution{University of Toronto}
}
\author{Timothy R. Langlois}
\orcid{0000-0002-5043-8698} 
\affiliation{%
  \institution{Adobe Research}
  \streetaddress{801 N 34th St}
  \city{Seattle}
  \state{WA}
  \postcode{98103}
}
\email{tlangloi@adobe.com}

\author{Yijiang Huang}
\affiliation{%
 \institution{MIT}
}

\author{Caitlin Mueller}
\affiliation{%
 \institution{MIT}
}

\author{Wojciech Matusik}
\affiliation{%
 \institution{MIT CSAIL}
}

\author{Ariel Shamir}
\affiliation{%
 \institution{The Interdisciplinary Center}
}

\author{Karan Singh}
\affiliation{%
 \institution{University of Toronto}
}

\author{David I.W. Levin}
\affiliation{%
 \institution{University of Toronto}
}

\renewcommand\shortauthors{Arora, R. et al}

\begin{abstract}
We present the first algorithm for designing volumetric Michell Trusses. Our method uses a parametrization approach to generate trusses made of structural elements aligned with the primary direction of an object's stress field. Such trusses exhibit high strength-to-weight ratios.
We demonstrate the structural robustness of our designs via a posteriori physical simulation.
We believe our algorithm serves as an important complement to existing structural optimization tools and as a novel standalone design tool itself. 
\end{abstract}

%
%


%
%


\begin{teaserfigure}
  \centering
  \includegraphics[width=\textwidth]{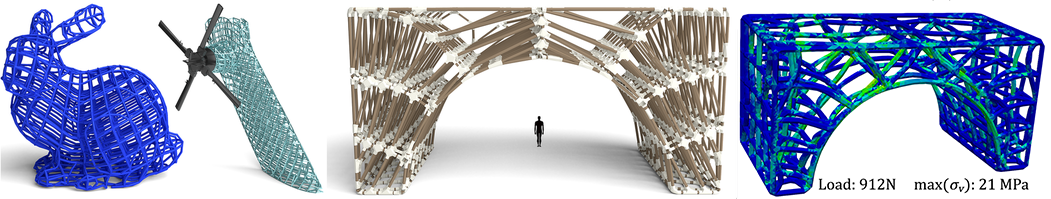}
  \caption{We generate stress-aligned 3D trusses which are 
  structurally-sound and lightweight. The structures 
  produced by our method consist of families of smooth, continuous curves tracing
  stress lines. 
  Representative examples demonstrating the versatility of our approach (from  
  left to right): the Stanford bunny, the rear pylon of a single engine helicopter, a bridge made of wooden beams and a simulation showing that a miniature (20cm wide) plastic bridge could support a 93 kg (205 lbs) person.
  \label{fig:teaser}}
\end{teaserfigure}

\maketitle
\thispagestyle{empty}

\section{Introduction}\label{sec:intro}
It is sometimes said that the primary objective of engineering is to develop 
the stiffest possible structure by using the least amount of material \cite{Doubrovski-2011}.
\ud{This guiding principle can be seen in many everyday structures such as bridges and stadiums. Strength-to-weight trade-off is naturally expressed as an optimization problem and its solution has become a foundational challenge in mathematics, computer science and engineering.}

\begin{figure*}
\includegraphics[width=\textwidth]{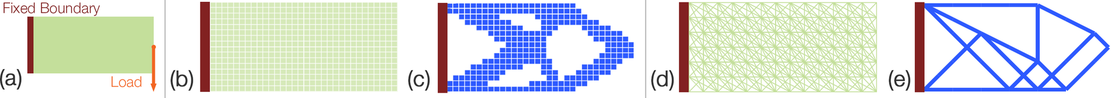}
\caption{Optimization of a cantilever beam (a) \ud{using voxel-based continuum-method (c) } and a
 ground structure method (e). \ud{Voxel-based continuum-methods}  optimize for material placement (c) 
 on the voxelized domain (b), while the ground structure method over samples the 
 domain with truss members (d) and solves for the optimum set of members (e).
\label{fig:topopt}}
\end{figure*}

\ud{Almost all structural optimization algorithms discretize the material distribution within the structure and then attempt to sparsify this distribution (see~\autoref{fig:topopt}). The nature of this discretization, be it voxels, level-sets or trusses, gives birth to the specific optimization and algorithm applied.} 

Unfortunately, all of these methods have inherent limitations, rooted in the requirement of an
overprescribed set of design variables (either voxels or bars) as initialization. The voxel grids of traditional Topology Optimization often require the use of additional regularization 
terms in the optimization objective in order to avoid ``checker-boarding'' artifacts, level-sets can require additional foliation terms to generate topology change and truss-based methods require an appropriate set of input trusses be specified. 

\begin{figure}[htb]
\includegraphics[width=\columnwidth]{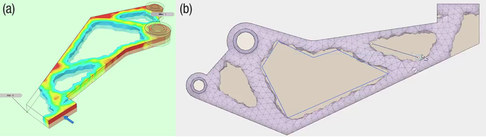}
\caption{Topology optimization results can be challenging to fabricate using non-additive manufacturing techniques. Even when using state of the art commercial tools for topology optimization such as Fusion 360~\cite{fusion360} (a), users have to manually trace over the results (b) to produce a fabricable geometry. \textcopyright\ Autodesk Sustainability Workshop \mbox{\url{https://youtu.be/lyTULzvHhXw}}.
\label{fig:tracing}}
\end{figure}

\ud{Even when such difficulties can be overcome, these methods (by construction) lack any global notion of object topology or geometry.  This can have implications down stream during the design process wherein an architect or engineer may wish to make small changes to the design, such as constraining certain points, deleting structural members or consistently resizing elements. The importance of these operations, coupled with the difficulty in performing them on standard topology optimized output, has led practitioners to employ frustratingly manual solutions such as tracing over optimized results to produce a final fabricable object (\autoref{fig:tracing}).}

In this paper, we take a different approach to the generation of light, and strong
structures. Instead of starting with an overprescribed solution and sparsifying it, we 
use one-dimensional cylindrical structural elements to define a truss, and formulate its design as 
a fitting problem for these elements. 
Michell~\cite{Michell:1904:Limits} laid the
foundations for creating such trusses by proving that for a given material budget, all elements of 
the optimal (stiffest) truss must follow paths of maximum
strain. Structures which fulfill this property are
called \emph{Michell Trusses}. Hence, by aligning the individual elements 
with the principal stress directions of an object's stress tensor field, a structurally sound design can 
be created without needing to fill the entire shape volume with material (and later sparsifying it).




Michell Trusses can be difficult to design for all but the
simplest geometries and loading conditions. While a number of recent
works have shown how to computationally design Michell Trusses for 2D domains or on 
height-fields, an algorithm for generating Michell Trusses for arbitrary loads acting
on three dimensional shapes has remained elusive until now.

We present the first algorithm to design truss structures following Michell Truss
principles inside arbitrary 3D domains.  Rather than optimizing an initial guess, we treat truss
optimization as a fitting problem. 
Our method
requires only a single solve of the static equilibrium equations to compute a
continuous stress field. We then use a novel parametrization method to
produce a graph of a prescribed resolution where each graph edge is as aligned as
possible with the underlying stress tensor field. Our method avoids many of the
difficulties of previous methods, its initialization is trivial, and we require
no additional regularization terms to avoid high-frequency artifacts in our
results. 


\ud{
\paragraph{Contributions}
To summarize, the main contribution of our work is the first algorithm for generating 3D Michell Truss structures on complex 3D geometry.
In realization of this overarching goal, we make the following three technical contributions:}

\begin{itemize}
\item A method for extracting a volumetric, stress-aligned frame field.
\item A method for generating a volumetric texture parametrization with coordinate lines aligned with the frame field.
\item A method for extracting the truss structure from a volumetric texture.
\end{itemize}

We show results on various 2D and 3D examples and demonstrate the high
strength-to-weight ratio we achieve compared to na\"{i}ve truss layouts.
\section{Related Work}\label{sec:related}
Structural optimization is a classic problem in computational design,
fabrication and digital manufacturing. Methods exist to help designers identify
the absolute weakest parts of objects~\cite{Zhou:2013:WSA} or the weakest parts
under real world forces~\cite{Langlois:2016:Stopt}. Other methods attempt to
reinforce designs to improve their strength~\cite{Stava:2012:SRI} or find the
most stable orientation for 3D printing a design~\cite{Umetani:2013:CSA}.

In this paper, we focus on the problem of generating structurally sound objects
via optimal material placement. Algorithms for this task define optimality using some measure of an object's
strength---most often attempting to minimize an object's compliance under a
given load~\cite{Bendsoe:2009:Topology,Freund:2004:Truss} while satisfying
constraints on the amount of material used. 

\subsection{Voxel and Level-Set Optimization}
\ud{Algorithms for structural optimization can be differentiated based on their chosen discretization for the material distribution. One oft-used material discretization is a dense
voxel grid (though other mesh structures can be
used~\cite{GAIN2015411,HA20081447}).} This approach has recently been used to create extremely detailed designs such as bone-like
structures~\cite{Wu:2017:Infill} and even airplane wings~\cite{Aage:2017:Giga}. Variations on a theme include using image slice stacks to generate infill for 3D prints~\cite{MAO_2018}. 
While capable of generating a wide range of stable designs, these methods can be
difficult to control, requiring regularization to avoid non-physical
``checkerboarding'' artifacts (high frequency
patterns of solid and empty voxels) and disconnected components (which make the designs
un-manufacturable)~\cite{Schumacher:2015:MCE}. They also require a choice of density
cutoff which determines when a cell is considered empty or not. Voxel-based topology optimization is also unsuitable for many mission-critical engineering applications as it can create 
internal cavities and tiny internal features making surface inspection 
impossible~\cite{NASA:2014:NDE, AFRL:2014:NDE}.

\ud{Level-sets have recently become a popular material distribution representation, in part because they help avoid some of the artifacts of a discrete voxel representation~\cite{WANG:2003:LS}. Level-sets can represent smoother geometry and their use mitigates ``checkerboarding''.  However, such methods must rely on foliation terms to instigate topology change with the final outcome depending on the topology of the initial solution~\cite{ALLAIRE:2004:SO}.}

\ud{There have also been efforts to improve the performance of these types of high resolution material optimization methods. For instance, by assuming that the outer shape is fixed, Ulu et al.~\shortcite{Ulu:2017:LSD} show how 
to leverage model reduction to speed up internal structure optimization by reducing the 
number of variables.}

\subsection{Truss Optimization}
Truss-based optimization methods~\cite{Bendsoe1994,Freund:2004:Truss} are attractive for
their small number of design variables (compared to voxel-based methods) and
ease of manufacturing. Michell~\shortcite{Michell:1904:Limits} first discovered
that an optimal truss layout (in terms of strength-to-weight ratio) is given
when trusses are aligned with the principal stress directions induced by loading
conditions. \ud{Intuitively, this aligns elements with the directions of pure compression and tension minimizing stress due to bending}. In certain cases, it is possible to solve for this optimal layout
analytically~\cite{Jacot:2017:Strain}, but no analytical solution is known for
the general case, so the ground structure method
(GSM)~\cite{Dorn:1964:AutoOptimal,Zegard:2015:GGS} is used. Here, an initial
layout of a finite number of trusses is specified, and the radii and
connectivity of the trusses are optimized to minimize the total weight. The
traditional GSM formulation suffers from several problems:
\begin{enumerate*}[label=(\arabic*)]
\item an initial layout of node positions needs to be specified, which can limit the solution space;
\item it can yield self-intersecting beams;
\item it assumes that the cross-section of each truss member can be set independently, which makes large-scale manufacturing challenging.
\end{enumerate*}

Multiple heuristic methods such as particle swarm
optimization~\cite{Li:2009:HPS}, ant colony optimization~\cite{Kaveh:2008:Ant},
teaching-learning-based optimization~\cite{Camp:2014:MTL}, and genetic
algorithms~\cite{Kawamura:2002:Genetic} have been proposed to address the third
problem by limiting the cross-sections of the trusses to a small set. Mixed-integer 
programming techniques have been used to achieve global optima for this 
problem~\cite{Achtziger:2007:Truss,Stolpe:2003:Mixed,Rasmussen:2008:CutBranch}.
However, these methods were only demonstrated on small models. Jiang et
al.~\shortcite{Jiang:2017:DVO} recently demonstrated much larger examples by
dividing the mixed-integer problem into three subproblems that were solved
iteratively. This works well in practice, but does not guarantee a globally
optimal solution. The method also optimizes initial node positions and
connectivity to avoid self-intersections, but still requires an oversampled
initial mesh, the design of which remains challenging. 

\subsection{Stochastic and Spectral Approaches}
Martinez et al.~\shortcite{Martinez:2017:OKN} generate procedural
anisotropic ``foams'' by warping the local distance metric. They show that by
controlling this warping they can generate 3D printable metamaterials with
anisotropic mechanical properties, and that they can align their
metamaterials with the stress field resulting from a 2D topology optimized
structure. In contrast to this approach, our method operates on a larger, macroscopic
scale, does not require an initial topology optimization pass, and provides us
with much greater control of the resolution of the produced structure. We also show that our
method can produce fabricable output from 3D stress fields, not just 2D. Nguyen et al.~\cite{Nguyen:2012:Conformal} generate a conformal cell structure
from a predefined set of cells. However, the radii of trusses can still vary
uniquely, which makes manufacturing difficult. Wang et
al.~\shortcite{Wang:2013:CPO} generate skin-frame structures with a solid outer
shell and strut filled interior to reinforce objects for 3D printing. The method
requires an initial internal sampling of nodes; struts are filled in with an ANN
method, and an $\ell_0$ sparsity optimization is used to minimize the number of
struts.

There
have also been more user-centric algorithms proposed. For instance Zehnder et
al.~\shortcite{Zehnder:2016:DSO} propose an interactive design tool for
constructing ornamental curve networks on surfaces. They use a spectral approach
to determine structural stability, ensuring the design has no low energy
deformation modes.

\subsection{Michell Trusses}
\ud{Though some questions have been raised regarding the optimality of Michell Trusses~\cite{Sigmund:2016:OMS}}, \ud{they still play an important role in engineering design.} While the methods above try to 
approximate a truss layout via optimization, \ud{~\cite{Bendsoe1994}}, other methods more directly attempt to generate Michell layouts.
Tam et al.~\cite{Tam:2015:Stress,Tam:2017:AddMan} generate principal stress
lines directly by integrating the stress field, and develop a novel robotic arm
printer capable of printing along these lines directly. However, currently the
method only applies to 2.5D structures (i.e., structures that are 3D but only
need one layer of trusses, such as shells or membranes), and cannot handle 3D
volumetric cases.  Li
and Chen~\shortcite{Li:2010:Beam} begin with a (very simple) user provided beam
network which connects the contacts to the points of application of external
forces. Then, an iterative method subdivides this structure and better
approximates principal stress lines until the desired compliance/strain energy
is achieved. While motivational, this method only works for 2D structures. It is
also unclear if the user interaction is amenable to more complicated shapes or
contact/load configurations. \ud{Li et al.~\shortcite{Li:Rib:2018} produce rib like reinforcements aligned with principle stress directions but again, only for 2.5D structures.}

\subsection{Parameterization-Based Mesh Generation}
Our method replaces structural optimization with automatic mesh
generation and provides the first (to the authors' knowledge) algorithm for
computing Michell Trusses in arbitrary 3D shaped domains under arbitrary
loading conditions.

Our method is inspired by recent developments in hex and
quad meshing for 3D geometries (for instance
~\cite{Panozzo:2014:FFA,Nieser:2011:CubeCover}). These algorithms use prescribed
frame fields to align the gradients of a volumetric function such that a hex
mesh can be extracted. The general hex meshing problem is hard and still an
active area of research. \ud{None of the currently available algorithms satisfy the criteria necessary for solving our particular problem}.

The seminal paper,
CubeCover~\cite{Nieser:2011:CubeCover}, solves a generalized version of the
parametrization and mesh extraction problem we solve. However, their method must
introduce discrete optimization variables in order to compute a well aligned
frame field. \ud{We confirmed via personal communication that CubeCover does not generate approriate frame fields and thus ``it's impossible for the software in its current state to be used as stand-alone solution''~\cite{CubeCoverPC}}. 
\ud{Ray et al.~\shortcite{Ray:2016:PFF} and Solomon et
al.~\shortcite{Solomon:2017:BEO} tackle the issue of frame field generation by introducing functional representations of frames.} Ray et
al.\ align their frame field with a mesh boundary and smoothly interpolate into
the object volume. Solomon et al.\ also smoothly interpolate inside the
object volume using a boundary element based approach. \ud{However, our problem requires the opposite objective as we
care little for boundary alignment and much more for accurate alignment within
the mesh volume itself.} $L_p$-centroidal voronoi tessellation~\cite{Levy:2010:LPC} can generate
hex-dominant meshes that take a background anisotropy field into account.
However, it does not follow the anisotropy as closely as our method does.
Lyon et
al.~\shortcite{Lyon:2016:HRH} present a method for mesh extraction; that is,
given a parametrization on a tetrahedral mesh, they extract out a 3D-embedded
graph. However, their method requires that the input parametrization is
boundary-aligned. \ud{Again, our method requires good alignment in the interior of the object, not the boundary, making this method unsuitable.} Many of our results
such as a bridge or beam
(Figs.~\ref{fig:resultsSim},~\ref{fig:resultsSSFixed}) are naturally
stronger when trusses are not normal to the boundary. \ud{Unlike all of the methods above, ours is the first to generate global, structurally sound parameterizations.}

\section{Background and Preliminaries}\label{sec:preliminaries}
We begin by introducing the technical background necessary to understand our formulation,
 starting with an introduction to truss optimization.

\subsection{Truss Optimization}
\ud{A \emph{truss} is a structure consisting of a network of members, each of which
is under purely axial stress. Typically, the forces only act at the joints between these members,
known as the \emph{nodes} of a truss.} Given a domain $\Omega\subset\BR^d$, and boundary
conditions consisting of a set of static forces applied on the boundary and
anchoring parts of the boundary to fixed supports, truss optimization
is the problem of finding a structurally sound truss minimizing the volume of
material utilized. In general, this involves optimizing over three design parameters:
\begin{enumerate}[label=(\arabic*)]
\item the \emph{topology}, which specifies the connectivity of the truss members;
\item the \emph{geometry}, specifying the positions of the nodal points; and
\item the \emph{size}, which gives the cross-sectional area of the members.
\end{enumerate}
We use the term \emph{truss layout} to refer to the choice of topology and geometry 
of the truss. Mathematically, this is equivalent to a graph embedded in $\Omega$.

In the classical truss optimization formulation, known as the Ground Structure method,
 an a priori chosen set of
uniformly spaced nodal points and members cover the problem domain $\Omega$, forming
the so called \emph{ground structure}. The topology of the optimal truss is generated
by varying the cross-sectional areas of the members, allowing for zero areas which
effectively removes those members. Since the nodal points are assumed to be fixed, the
classical approach only solves for the topology and size parameters.

\subsection{Michell Truss Theory}

\begin{figure}
\centering
\includegraphics[width=\columnwidth]{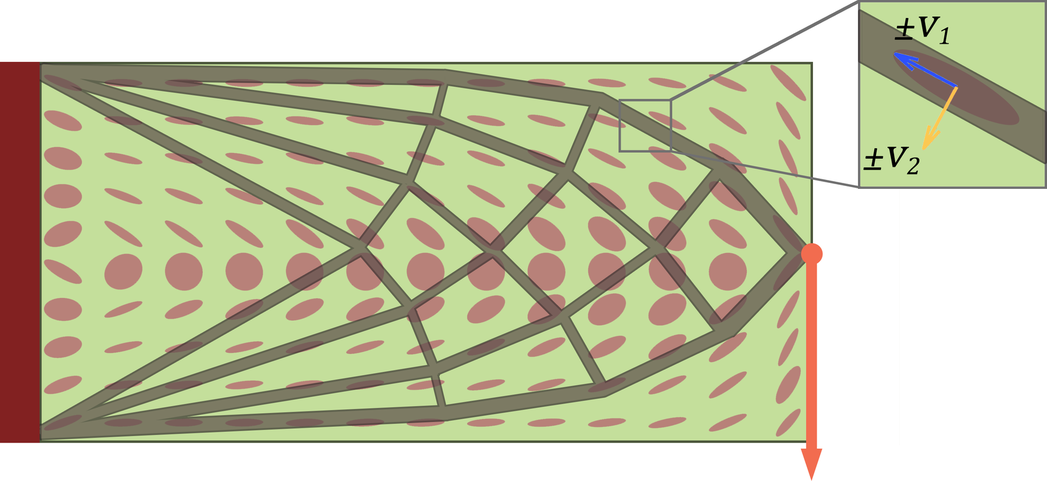}
\caption{Michell Truss members are aligned with the principal directions of the
underlying stress field. Here, we show the example of a cantilever. The ellipses 
visualize the stress tensor, and the optimal
Michell truss for a chosen discretization is overlaid on the problem domain.
Inset: Michell truss member aligned with a stress
eigenvector.
\label{fig:michellTruss}}
\end{figure}

Michell's theorem~\citet{Michell:1904:Limits} characterizes the fundamental
properties of the optimal truss structure, called the Michell truss, for the
problem defined above. The theorem states that the members of the optimal truss
structure lie along the \emph{principal directions} of the \emph{virtual} stress
field. The virtual stress field is defined as the stress induced by the given
external forces if the domain were to be uniformly filled with material, and principal
stress directions simply refer to the eigenvectors of the stress tensor matrix.
Owing to the continuity of the stress tensor field and the orthogonality of
eigenvectors of a Hermitian matrix, the principal stress directions form a set of families
of orthogonal curves called the principal stress lines. In the continuous setting, an 
optimal Michell Truss consists of an infinite set of infinitesimally small members tracing
these curves. Computationally, the optimal truss layout for the chosen
discretization consists of finite sized members approximating the principal stress
lines (see \autoref{fig:michellTruss}).

In the Ground Structure method, \ud{the approximation error is determined by the
density and connectivity of the initially chosen ground structure.}
Unfortunately, it is difficult to choose an appropriate discretization balancing
accuracy and computational time for complex domain geometries.

\section{Stress-Aligned Truss Network Generation}\label{sec:method}
\begin{figure*}[htp]
    \includegraphics[width=\textwidth]{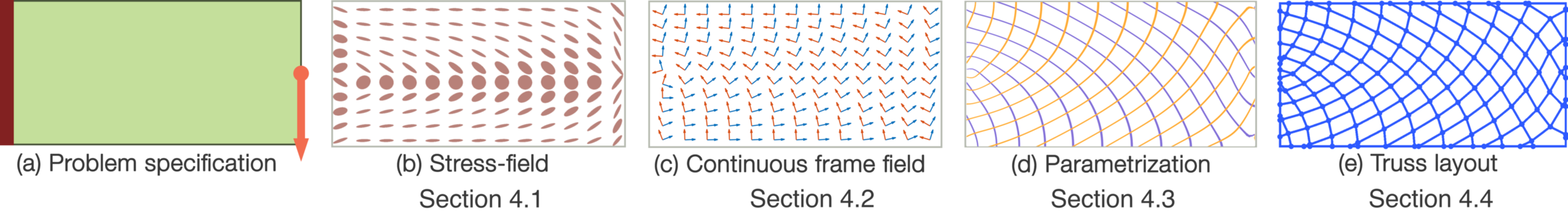}
    \caption{Overview of our method: (a) starting with a problem domain (green)
    with fixed points (red) and loads (orange) as boundary conditions, we first
    perform FEM analysis to compute the stress field (b). A continuous and
    smooth frame field (c) aligned with the principal stress directions is then
    computed. The components of this frame field are then used to compute a texture
    parametrization on the domain (d), whose isocurves (show in orange and
    purple) are traced to extract a stress-aligned truss structure (e).}
    \label{fig:overview}
\end{figure*}

We take a parametrization-based approach to generating stress aligned
trusses. Our algorithm
consists of four independent phases: (1) stress tensor field generation using
finite element analysis (FEA), (2) stress-aligned frame field fitting, (3) volumetric
texture parametrization, and (4) structural member extraction (Figure~\ref{fig:overview}). Optionally, we
can choose to sparsify a previously generated truss layout to reduce material use.

One might ask why follow such an approach given that robust and reliable hex
meshing for arbitrary geometries is, as yet, unsolved. Fortunately, our problem 
is more amenable to solution than that of general hex meshing as we have a volumetric 
tensor field to guide us. We also do not require hexahedral cells everywhere in our 
domain, or a boundary aligned  structure.

These
considerations eliminate some of hex-meshing's most aggravating difficulties,
allowing us to develop a flexible algorithm, which, as we will demonstrate, can
be applied to a wide variety of geometries.
In the remainder of this section, we will detail each step of our truss generation method.
\subsection{Finite Element Analysis}\label{sec:fem}
The first step of our method is to generate a stress tensor field for an input
geometry. We use standard \ud{linear elastic~\cite{Gould:1994:Elasticity}} finite element
analysis with tetrahedral discretization~\cite{Belytschko:2013:Nonlinear,Levin:2017:GAUSS}
for this task. Both Dirichlet and
Neumann boundary conditions are manually determined based on the expected loading
conditions of the given shape. We then compute the Cauchy stress tensor field
(for our elements, a single tensor per tetrahedron) for use in subsequent algorithmic 
stages. We use the same discretization for all steps of the method.

\ud{In the remaining sections, we refer to the continuous input domain as $\Omega\subset\BR^3$ and the tetrahedral discretization of the domain as the mesh $\cM = (\cV, \cT)$.}

\subsection{Stress Aligned Frame Field Generation}\label{sec:vectorField}

\setlength{\intextsep}{0pt}
\setlength{\columnsep}{0pt}
\begin{wrapfigure}[9]{r}{0.2\linewidth}
    \includegraphics[width=\linewidth]{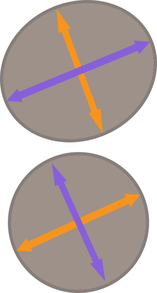}
\end{wrapfigure}
\setlength{\intextsep}{3pt}
\setlength{\columnsep}{10pt}

Na\"ively, a Cauchy Stress tensor field $\sigma\left(\bx\right)$ can be
interpreted as a frame field by representing each tensor by its three
eigenvectors. Because each $\sigma\left(\bx\right)$ is a Hermitian matrix, 
its eigenvectors are guaranteed to form an orthogonal basis.
However, such a frame field is almost certain to be non-smooth as the direction
of each eigenvector can be arbitrarily flipped or interchanged (see inset). Previous
algorithms for frame field generation~\cite{Nieser:2011:CubeCover} handle such
reflection symmetries by searching over all possible symmetric frame
configurations. This is effective but complicates optimization by introducing
discrete variables.

Rather than using discrete variables, we are inspired by methods which work with inherently
symmetric, functional representations of frame
fields~\cite{Ray:2016:PFF,Solomon:2017:BEO}. We are also influenced by
Levin et al.~\shortcite{Levin:2011:Muscle} which shows that by using the Rayleigh
Quotient~\cite{horn1990matrix} as an objective, one can produce vector
fields that smoothly align with the most locally anisotropic direction of a
tensor field. The key observation is that the tensor itself is a useful,
symmetry agnostic frame field representation and here we leverage this notion
and extend it to frame field fitting.

\ud{A 3D frame can be encoded using three unit vectors.
We define the notion of alignment of a single unit vector with the stress tensor using the square root of the absolute value of the Rayleigh Quotient, and introduce the notation $\|\cdot\|_M$, where $M\in\BR^{3\times 3}$. 
Now, for unit vectors $\bv $, $\|\bv\|_M = (|\bv^T M\bv|)^{1/2}$ is maximized when $\bv$ is aligned with the primary eigenvector of $M$.} 

\ud{Let the eigendecomposition of $\sigma$ be given by $\sigma = Q\Lambda Q^T$, where $\Lambda$ is a diagonal matrix whose diagonal elements are the sorted (in decreasing order) eigenvalues of $\sigma$. We define
\begin{equation}
	\sigma_+ = Q \lambda Q^T,
	\label{eq:pdSigma}
\end{equation}
where the operation $\lambda = (|\Lambda_{ij}|)$ returns a matrix with entries of $\Lambda$ replaced by their absolute values.}

\begin{figure}[t!]
    \includegraphics[width=\columnwidth]{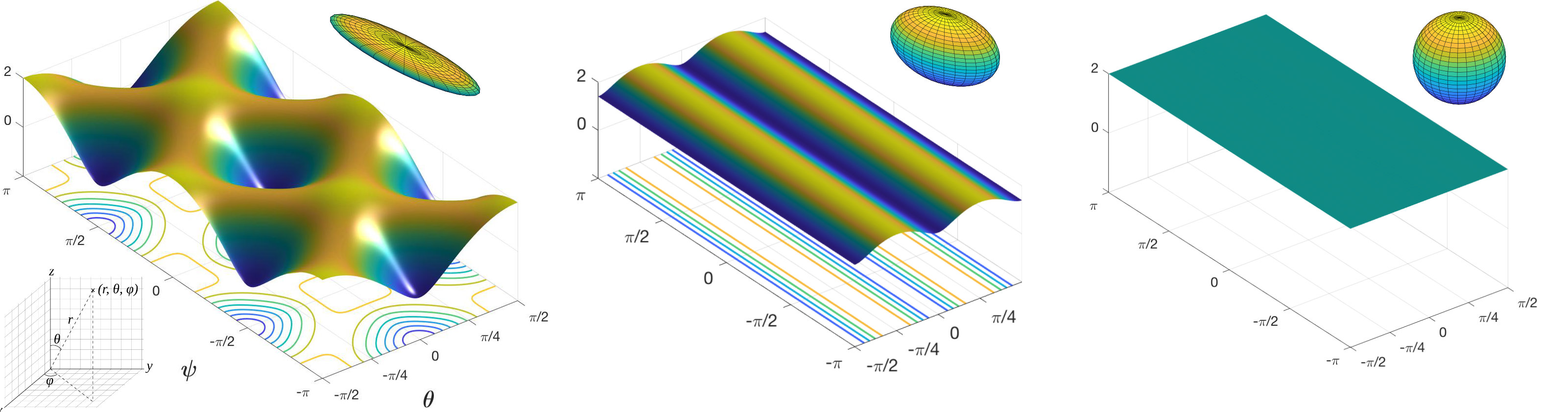}
    \caption{Our cost function has identical local minima corresponding to any
    orthogonal transformation that aligns a frame with the second and third
    eigenvectors of a tensor. This behavior is consistent for tensors with 3 
    (left) 2 (center) and 1 (right) distinct eigenvalues.
    \label{fig:cost}}
\end{figure}

We are now ready to define alignment of frames and stress tensors.
We define a ``good'' fit between a frame and a stress tensor as one where the
first axis of the frame is aligned with the primary eigenvector of the stress
tensor and the other two axes are aligned with the second and third eigenvectors
(though it does not matter which aligns with which). To this end we define the
following frame-tensor matching function:
\ud{
\begin{equation}
    E^i_{data}\left(R=(\br_1, \br_2, \br_3)\right) = \|\br_2\|_{\sigma_+^i} + \|\br_3\|_{\sigma_+^i},
    \label{eq:data}
\end{equation}
where $\sigma^i\in\BR^{3\times3}$ is the stress tensor of the
$i^{th}$ tetrahedron in our mesh $\cM$ (we use the simulation discretization for the
fitting stage of our method as well), $R$ is our frame with $\br_j\in\BR^3$ its $j^{th}$
direction vector. The positive-definite matrix $\sigma_+^i$ is defined according to~\autoref{eq:pdSigma}.}
This cost function has a set of identical minima at
every frame alignment which satisfies our criteria~(\autoref{fig:cost}).
\ud{In order to improve the numerical behavior of this energy function, we rescale $\lambda'$ so that the eigenvalues lie in the range $[1, 30]$ (chosen experimentally). Note that this rescaling alters neither the eigenvectors, nor the ordering or multiplicity of the eigenvalues of $M$, making $\|\cdot\|_M$ a true norm.}

Next we need a method for disambiguating the local minima in \autoref{eq:data}.
Typically this is done combinatorially, but here we follow the approach of
Solomon et al.~\shortcite{Solomon:2017:BEO} and instead use a smoothness energy
to produce a well-fitted, consistently aligned frame field. Solomon et
al.~represent rotations using canonical axis functions and use a standard
Laplacian smoothing term. While we borrow their smoothing energy, we
cannot use their frame representation as it requires an extra, approximate
projection to produce proper, orthogonal frames. In our problem, alignment with the data is
critical, and introducing error via such a projection is not acceptable.

Instead we represent a frame at the centroid of a tetrahedron using rotation
matrices, parameterized via the matrix exponential, $R^i =
\expm\left(\sum_{j=1}^4\left[\omega^j\right]\right)\in\BR^{3\times 3}$. Here,
$\omega^j\in\BR^3$ are angular velocity vectors at the vertices of each
tetrahedron and the $\left[\cdot\right]$ operator computes the cross product matrix from an input vector. This allows us to define a smoothness
energy in the following manner:
\begin{equation}
    E_{smooth}\left(\mathbf{\omega}\right)	= \frac{1}{2}\mathbf{\omega}^TL\mathbf{\omega} \ud{\ +\ \frac12 \mathbf{\omega}^T\mathbf{\omega}},
    \label{eq:smooth}
\end{equation} where $\mathbf{\omega}$ is the stacked vector of all per-vertex $\omega^i$'s,
and $L$ is a block diagonal cotangent Laplacian matrix. \ud{The second term regularizes $\mathbf{\omega}$ and prevents it from taking arbitrarily large values. In practice we found this helped with the stability of the line search of our optimization scheme (fmincon~\cite{matlab})}

Initially we attempted to perform frame fitting using a weighted sum of \autoref{eq:data} and \autoref{eq:smooth}:
\begin{align}
    E_\alpha(\mathbf{\omega}) &= \sum_{i=1}^N E_{data}^i\left(\bm{r}_1\left(\omega\right), \bm{r}_2\left(\omega\right),\bm{r}_3\left(\omega\right)\right) + \alpha E_{smooth}\left(\mathbf{\omega}\right)\\
    \mathbf{\omega}^* &= \underset{\mathbf{\omega}}{\argmin}\ \ E_\alpha(\mathbf{\omega})
    \label{eq:finalCost}
\end{align}
where $N = |\cT|$ is the number of tetrahedra in $\cM$,
$\alpha$ is a scalar weight, and $R^i = \left(\bm{r}_1,\bm{r}_2, \bm{r}_3\right)$.
Minimizing this cost function, using an
L-BFGS Hessian approximation, revealed issues in choosing an appropriate
$\alpha$. To alleviate this problem, we again lean on the fact that our sole
concern is minimizing the data term. The only purpose of the smoothness energy
is to help us choose an appropriate local minima to descend into. To this end,
our final fitting algorithm (Algorithm~\ref{alg:frameFitting}) is Augmented
Lagrangian-esque~\cite{nocedal2006numerical} in that we repeatedly
minimize~\autoref{eq:finalCost} with increasingly smaller $\alpha$ until the
data cost stops decreasing. In practice our termination criteria was not
complex: a fixed \ud{thirty} iterations after each of which $\alpha$ was \ud{reduced to $(2/3)\alpha$}. This
proved to be more than enough for all our examples and yielded excellent
results.
There is a minor implementation
detail which arises when using matrix exponentials as a parametrization of
rotation matrices---their gradient is undefined at $\mathbf{\omega} =
\mathbf{0}$. We avoid this problem by perturbing any 0 length angular velocity
vector by the square root of machine epsilon (a standard work-around). Other terms in the gradient
ensure good numerical behavior near the singularity.

\begin{algorithm}
$\mathbf{\omega} \leftarrow \mathbf{0}$\;
\ud{$\alpha \leftarrow 10|\cT|$\;}
\Repeat{convergence}{
	\tcc{Initialize L-BFGS with previous estimate of $\omega$ to solve \eqref{eq:finalCost}}
	$\mathbf{\omega} \leftarrow$ L-BFGS($E_\alpha$, $\omega$)\;
	\ud{$\alpha \leftarrow \frac23\alpha$\;}
}

\KwRet{$\omega$}\;
\caption{Iterative method for computing a stress-aligned frame field.\label{alg:frameFitting}}
\end{algorithm}
%

\subsection{Induced Parametrization Computation}\label{sec:parametrization}
We use our smooth, data-aligned frame field to compute a stress-aligned
parametrization from which we will create our Michell Truss. We define
$\Omega \subset \BR^3$ as the world space that our object occupies and
$\mathbf{u} \in \BR^3$ as a volumetric texture domain. We chose our
structural members to lie along the coordinate lines of $\mathbf{u}$ and seek to
find a parametrization $\mathbf{u}=\phi\left(\mathbf{x}\right):\Omega\rightarrow\BR^3$ that aligns
these coordinate lines with our frame field.
Formally we seek a $\phi\left(\mathbf{x}\right)$ such that
\setlength{\arraycolsep}{3pt}
\begin{equation}
    \begin{aligned}
        \frac{\partial \phi}{\partial \mathbf{x}}\mathbf{r_1} &= \begin{bmatrix} 1 & 0 & 0\end{bmatrix}^T& \\
        \frac{\partial \phi}{\partial \mathbf{x}}\mathbf{r_2} &= \begin{bmatrix} 0 & 1 & 0\end{bmatrix}^T& \\
        \frac{\partial \phi}{\partial \mathbf{x}}\mathbf{r_3} &= \begin{bmatrix} 0 & 0 & 1\end{bmatrix}^T&, \\
    \end{aligned}
    \label{eq:fitCriteria}
\end{equation} at the center of each tetrahedron in our mesh.
\setlength{\arraycolsep}{6pt}

We can write these requirements as a linear system of equations by constructing the discrete directional gradient operator for each tetrahedron in our mesh:
\begin{equation}
    G^i\left(\mathbf{v}\right) =
    \begin{bmatrix}
        \mathbf{v}^i_x\cdot G^i_x + \mathbf{v}^i_y\cdot G^i_y + \mathbf{v}^i_z\cdot G^i_z
    \end{bmatrix},
\end{equation} where $G_x$, $G_y$ and $G_z$ are the discrete gradient operators of
our tetrahedral mesh, $\mathbf{v}\in \BR^3$ is the direction in which
the derivative is to be measured (at the centroid of a tetrahedron) and $i$
indexes our tetrahedra. We can assemble these local directional derivative
operators into global matrices to produce the global operator $G\left(\mathbf{v}\right)$.

We proceed by constructing three directional derivative operators, one for each frame director 
\begin{equation}
    \begin{aligned}
        \bG_1 &= G\left(\mathbf{r_1}\right)& \\
        \bG_2 &= G\left(\mathbf{r_2}\right)& \\
        \bG_3 &= G\left(\mathbf{r_3}\right)&. \\
    \end{aligned}
    \label{eq:discreteGs}
\end{equation}
\ud{The discrete version of~\autoref{eq:fitCriteria} can then be stated as the following constrained minimization problem.}
\setlength{\arraycolsep}{3pt}
\ud{
\begin{equation}
	 \phi^*=\underset{\phi}{\argmin}\ \left\lVert
    \begin{bmatrix}
        \bG_1 & \mathbf{0} & \mathbf{0} \\
        \mathbf{0} & \bG_2  & \mathbf{0} \\
        \mathbf{0} & \mathbf{0} & \bG_3
    \end{bmatrix} \mathbf{\phi}
    -
    \begin{bmatrix}
        \mathbf{1}\\
        \mathbf{1}\\
        \mathbf{1}
    \end{bmatrix}
    \right\rVert^2_2,\\
    \label{eq:parametrizeCon1}
\end{equation}
\begin{equation}
    \text{s.t.}\
    \begin{bmatrix}
        \mathbf{0} & \bG_1  & \mathbf{0} \\
        \mathbf{0} & \mathbf{0} & \bG_1\\
        \bG_2 & \mathbf{0} & \mathbf{0} \\
        \mathbf{0} & \mathbf{0} & \bG_2 \\
        \bG_3 & \mathbf{0} & \mathbf{0} \\
        \mathbf{0} & \bG_3  & \mathbf{0}
    \end{bmatrix} \mathbf{\phi}
   =
    \begin{bmatrix}
        \mathbf{0}\\
        \mathbf{0}\\
        \mathbf{0}\\
        \mathbf{0}\\
        \mathbf{0}\\
        \mathbf{0}
    \end{bmatrix} \text{, and }\hspace{0.3cm}
    \begin{bmatrix}
        \bG_1 & \mathbf{0} & \mathbf{0} \\
        \mathbf{0} & \bG_2  & \mathbf{0} \\
        \mathbf{0} & \mathbf{0} & \bG_3
    \end{bmatrix} \mathbf{\phi}
   >
    \begin{bmatrix}
        \mathbf{0}\\
        \mathbf{0}\\
        \mathbf{0}
    \end{bmatrix}.
	\label{eq:parametrizeCon2}
\end{equation}
}
\ud{That is, we constrain the gradients of the parametrization to follow the frame field (eq.~\ref{eq:parametrizeCon2}),
while optimizing for a regular, equi-spaced, solution (eq.~\ref{eq:parametrizeCon1}).
Unfortunately, the above formulation turns out to be infeasible for most of our test cases.
A counting argument suggests a probable cause: our target solution has $3|\cV|$ variables, but we apply $9|\cT|$ constraints.
Since, typically $|\cT| \gg |\cV|$ for manifold meshes (with low genus), the problem is likely to be overconstrained.}

\begin{figure}[H]
\centering
\includegraphics[width=\columnwidth]{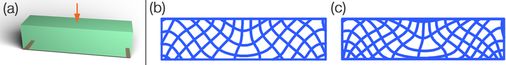}
\caption{Varying $\beta$ to control truss spacing for the bar with fixed ends problem:
the problem domain and boundary conditions (a), a slice of the truss extracted by setting
$\beta=1$, thus preferring regular spacing over orthogonality of the parametrization (b),
and the same slice with $\beta=0.1$, which favours parameter orthogonality (c).
\label{fig:frameFitBeta}}
\end{figure}

Therefore, we reformulate Equations~\ref{eq:parametrizeCon1}--\ref{eq:parametrizeCon2} into an unconstrained weighted quadratic minimization.

\begin{equation}
\begin{split}
    \phi^*=\underset{\phi}{\argmin}\ \beta \left\lVert
    \begin{bmatrix}
        \bG_1 & \mathbf{0} & \mathbf{0} \\
        \mathbf{0} & \bG_2  & \mathbf{0} \\
        \mathbf{0} & \mathbf{0} & \bG_3
    \end{bmatrix} \mathbf{\phi}
    -
    \begin{bmatrix}
        \mathbf{1}\\
        \mathbf{1}\\
        \mathbf{1}
    \end{bmatrix}
    \right\rVert^2_2	\ldots \\+
    \left\lVert
    \begin{bmatrix}
        \mathbf{0} & \bG_1  & \mathbf{0} \\
        \mathbf{0} & \mathbf{0} & \bG_1\\
        \bG_2 & \mathbf{0} & \mathbf{0} \\
        \mathbf{0} & \mathbf{0} & \bG_2 \\
        \bG_3 & \mathbf{0} & \mathbf{0} \\
        \mathbf{0} & \bG_3  & \mathbf{0}
    \end{bmatrix} \mathbf{\phi}
    -
    \begin{bmatrix}
        \mathbf{0}\\
        \mathbf{0}\\
        \mathbf{0}\\
        \mathbf{0}\\
        \mathbf{0}\\
        \mathbf{0}
    \end{bmatrix}
    \right\rVert^2_2
    \end{split}
\end{equation}
\setlength{\arraycolsep}{6pt}
where the parameter $\beta$ provides user control over the regularity of the truss spacing.
\autoref{fig:frameFitBeta} shows how the value of $\beta$ influences the solution.

\subsection{Truss Layout Extraction}\label{sec:layoutExtract}


In the final step of our algorithm, the parametrization is used to extract the
truss layout as a graph embedded in $\Omega$. In order to avoid confusion with the vertices
and edges of the input geometry, we exclusively use the words \textit{nodes} and
\textit{elements} to refer to the vertices and edges of the graph extracted from
the parametrization. Similar to parametrization based approaches for hex
meshing~\cite{Nieser:2011:CubeCover, Lyon:2016:HRH}, our aim is to trace the
integer isocurves of the parametrization. That is, we want the nodes $\cN$ of
the extracted graph to be the points mapped to integers, i.e.,
\begin{equation}
    \cN = \{\bx\in\Omega\ |\ \phi(\bx) \in \BZ^3\},
    \label{eq:nodes}
\end{equation}
and the elements $\cE$ connect adjacent points on an integer grid, i.e.,
\begin{equation}
    \cE = \{\{\bx,\by\}\ |\ \bx,\by\in\cN, \ \phi(\bx)-\phi(\by)\in\{\be_1,\be_2,\be_3\}\},
    \label{eq:elements}
\end{equation}
where $\be_i$'s are the standard basis vectors.

Note that only the gradient directions of our parametrization have any physical
meaning, and therefore, applying an arbitrary translation and/or scale
to the parametrization essentially keeps the physical information intact. In
order to provide user control over the density of the extracted truss, we first
translate and scale $\phi$ to normalize it to the range $[0, 1]$, and then
scale by a user-defined ``resolution'' parameter $\rho$. We refer to
this translated and scaled parametrization as
$\tphi=(\tphi_1, \tphi_2, \tphi_3)$.

As noted earlier, existing work on hex-meshing assumes that the parametrization
is defined such that for all points on the domain boundary, at least one of the
parameter values is an integer. Since this is not true for our parametrizations,
we also have to include additional nodes on the boundary, along with elements
connecting these to each other and to the internal nodes. We defer the
discussion of the special considerations for the boundary for now, and describe
our algorithm for tracing the internal elements first.

For ease of exposition, we start by describing a 2D truss layout extraction
algorithm. Recall that our stress fields do not have singularities in the interior of
the domain since forces are only applied at the boundary. This implies that
the principal stress lines must end at the boundary, and cannot end abruptly or
form closed surfaces inside the domain. Since the previous steps of the algorithm
ensure that the isosurfaces of $\tphi$ follow the principal
stress lines, we make the assumption that they do not form internal closed loops
as well. Therefore, for tracing these isocurves, we
start at their end points on
the boundary, and trace until we hit the boundary again.

\def\arraystretch{1.2}
\begin{center}
\begin{table}
	\begin{tabularx}{\columnwidth}{|p{0.75cm} | p{7.08cm}|}
	\hline
	\multicolumn{2}{|c|}{\textbf{Both 2D and 3D}}\\
	\hline
	$\partial\Omega$ & Boundary of the input domain\\
		$\partial\cM$ & The simplicial 2-complex bounding $\cM$\\
	$\cN$ & Points on the integer grid defined by $\tphi$ (the set of nodes of the truss layout)\\
	$\cE$ & The set of elements of the truss layout\\
	\hline
	\multicolumn{2}{|c|}{\textbf{2D}}\\
	\hline
	$\gamma_i$ & An integer isocurve of $\tphi_i$\\
	$\Gamma_i$ & The set of all $\gamma_i$\\
	$\cN_{Ei}$ & Intersection points b/w curves in $\Gamma_i$ and all edges of $\cM$\\
	$\cN_{E}$ & Disjoint union of $\cN_{E1}$ and $\cN_{E2}$\\
   $\cN_{e}$ & Points in $\cN_E$ lying on a particular edge $e$\\
	$\cN_{f,\gamma_i}$ & Points on the integer grid, lying on the intersection b/w $\gamma_i$ and a particular face $f$\\
	$\cN_{F}$ & Union of $\cN_{f,\gamma_1}$ and $\cN_{f,\gamma_2}$ over all faces of $\cM$\\
	$\cE_i$ & Set of all elements tracing integer isocurves of $\tphi_i$\\
	\hline
	\multicolumn{2}{|c|}{\textbf{3D}}\\
	\hline
	$\mathcal{S}_i$ & An integer isosurface of $\tphi_i$\\
	$\gamma_{ij}$ & An integer isocurve of $(\tphi_i, \tphi_j)$\\
	$\cN_{F}$ & Points of intersections between all $\{\gamma_{ij}\}$ and faces of $\cM$\\
	$\cN_{E}$ & Points of intersections between all $\{\gamma_{ij}\}$ and edges of $\partial\cM$\\
	\hline
	\end{tabularx}
	\label{tbl:notationTrussExtract}
	\ud{\caption{Notation for the truss layout extraction procedure described in~\autoref{sec:layoutExtract}.}}
\end{table}
\end{center}
\def\arraystretch{1}

We use $\Gamma_i$ to refer
 to the set of end-to-end integer isocurves of $\tphi_i$, and $\gamma_i$
 to refer to an arbitrary curve from this set. Note
that $\tphi$ is a piecewise linear field stored at the vertices, and
its value at arbitrary $\bx\in\Omega$ can be found using Barycentric
interpolation. However, Barycentric interpolation on a triangle is equivalent to
linear interpolation along edges followed by interpolating across a line
segment between two edges. We utilize this series of linear interpolations to
trace out the integer isocurves of our parametrization.

\begin{figure}[htb]
\centering
\includegraphics[width=\columnwidth]{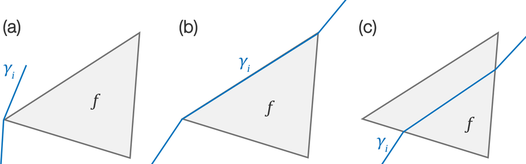}
\caption{An integer isocurve $\gamma_i$ can intersect a face $f$ either on a single 
vertex (a), on an edge (b), or go through its interior (c). We perturb the 
parametrization by an infinitesimal amount to eliminate the first two cases.
\label{fig:triangleIsocurve}}
\end{figure}

In order to make this approach work, we require that an isocurve and a face
intersect in exactly two points (or do not intersect at all). The only excluded cases are when an isocurve just touches a face at a single
vertex, or is aligned with one of the edges (\autoref{fig:triangleIsocurve}). 
While we never encountered these
cases with our paramterizations, we can easily eliminate the theoretical
possibility as well. We find the parameter values on vertices which are close to
integers up to machine precision, and translate them by $-\epsilon$, where
$\epsilon$ is a small positive number (we choose $10^{-7}$). If the vertex also
has the minimum parameter value in its 1-ring neighbourhood, we translate by
$+\epsilon$ instead, ensuring that we do not remove part of an integer isocurve.
This ensures that no $\gamma_i$ passes through a vertex, eliminating both the
problematic cases.

\begin{figure}[htb]
\centering
\includegraphics[width=\columnwidth]{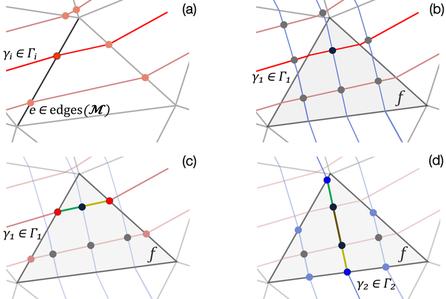}
\caption{In 2D, the truss extraction process begins by finding all points of intersections between
integer isocurves of $\tphi$ and edges of the input mesh (a). Then, for each face of the mesh $f$, 
and an integer isocurve of $\tphi_1$ intersecting with it, points mapped to the integer grid are located (b).
All such points form the set of nodes $\cN$ for the truss.
Finally, the linear section of each $\tphi_1$ isocurve is cut along these points to form the elements $\cE_1$
for the output truss (c), followed by a similar discretization and tracing process for $\tphi_2$ isolines to
form $\cE_2$ (d). The union of $\cE_1$ and $\cE_2$ is the set of elements $\cE$ of the truss.
\label{fig:trussExplain}}
\end{figure}

Starting from the parameter values stored at the vertices, we use
linear interpolation along edges to find the intersections of curves from $\Gamma_1$ 
and $\Gamma_2$ with the edges to form the sets of nodes $\cN_{E1}$ and 
$\cN_{E2}$, respectively (\autoref{fig:trussExplain}a). These nodes are then used to find nodes in the interior
of faces, and their connectivity, as described below.

Consider a
face $f$ and an isocurve $\gamma_1\in\Gamma_1$ intersecting with it. Let $\bx_0, \bx_1
\in \cN_{E1}$ be the end points of the line of intersection. We linearly
interpolate the values of $\tphi_1$ on $\bx_0$ and $\bx_1$ to find
the points of intersection of this isoline with all $\gamma_2\in\Gamma_2$:
\begin{equation}
    \cN_{f,\gamma_1}=\left\{\by\in f\cap\gamma_1\ |\
    \tphi_2(\by)\in\BZ\cap\left(\tphi_2(\bx_0),
    \tphi_2(\bx_1)\right)\right\},
    \label{eq:nodesInFace2D}
\end{equation}
where $\tphi_2(\bx_0) \le \tphi_2(\bx_1)$ \emph{wlog}. $\cN_{f,\gamma_1}$ is then
sorted by $\tphi_2$, the two extrema are connected to $\bx_0$ and $\bx_1$,
and consecutive points in the ordered set are connected to each other. Doing
this for all admissible pairs $(f, \gamma_1)$ gives the set $\cN_F$ of nodes
lying on the intersections between all pairs $(\gamma_1, \gamma_2)$, and the set of
elements $\cE_1$ tracing all $\gamma_1\in\Gamma_1$ (\autoref{fig:trussExplain}b--c).

Then, for each pair of intersecting face and $\gamma_2$, we search for the
nodes among $\cN_F$ lying on the intersection. The nodes lying on each
$\gamma_2$ are then connected to form the set of elements $\cE_2$ (\autoref{fig:trussExplain}d). Define
$\cN_E$ to be the disjoint union of $\cN_{E1}$ and $\cN_{E2}$, and
$\cN_{\partial\Omega}$ to be the subset of $\cN_E$ restricted to nodes lying on
the boundary. In the final step of the algorithm, we insert all nodes from
$\cN_{\partial\Omega}$ into a queue and trace out the integer isocurves
emanating from them, going through the faces it intersects until we hit the
boundary again. For each traced curve, we remove its endpoints from the queue.

\subsubsection{3D Truss Layouts}
\label{sec:layoutExtract3D}
In 3D we use  
$\mathcal{S}_i$ to denote an arbitrary end-to-end integer isosurface of 
$\tphi_i$, and $\gamma_{ij}$ to denote an arbitrary end-to-end 
curve where both $\tphi_i$ and
$\tphi_j$ are constant integers. After 
perturbing the parametrization, we compute the points of intersection of
isosurfaces of each of the three parameters with the edges.
Then, we compute the intersection points of each $\gamma_{ij}$ with all the faces of the mesh.
Finally, we linearly interpolate $\tphi_k (k\ne i,j)$ along these
isolines in each tet, and compute the intersections with all $\mathcal{S}_k$ to find
the elements of the truss.
Note that while the perturbation does not guarantee that every $\gamma_{ij}$
passes through the interior of all tets---it may just touch it at a face---we never
encountered this case in practice.

\subsubsection{Handling the boundary}
\label{sec:layoutExtractBoundary}
The described procedure already traces out
the nodes on the boundary, as well the elements which touch it. \ud{In 2D, we can
add the boundary by tracing the elements on each mesh edge individually---for an
edge $e=(\bx_0, \bx_1)$, sort all the points in $\{\bx_0, \bx_1\}\sqcup\cN_e$ by $\tphi_1$ (or equivalently, by $\tphi_2$) and connect adjacent points in the sorted list.}

\ud{In 3D, we need to trace the intersection of each $\mathcal{S}_i$ with the boundary $\partial\cM$, which comes down to performing the full 2D truss extraction procedure for each pair of parameters $\{i, j\} \in {3\choose 2}$ on the triangle mesh $\partial\cM$.}

\subsubsection{Simplification}
\label{layoutExtractSimplify}
Our extraction procedure can result in many nodes
which are geometric duplicates of each other, or are very close to each
other. However, we take care to get the correct graph topology in $\cE$ so that
such duplicates can be easily removed by performing edge contraction operations
on the graph with a small element length threshold. Optionally, we can
contract elements until we have no nodes from $\cN_F$ left ($\cN_E$ in 2D),
except those on the boundary. This gives us the graph similar to that in
Equations~\eqref{eq:nodes}--\eqref{eq:elements}, but with additional nodes and 
elements on the boundary.

We can simplify the boundary elements as well by contracting elements containing the
nodes in $\cN_E$ (boundary vertices in 2D). We perform both these steps for all
our results, but we do not remove boundary nodes which lie on feature edges
(feature vertices in 2D). These features are currently selected using dihedral
angles with a selection threshold of $\cos^{-1}(0.9)$ (approx. 25\textdegree),
but one could easily plug in user-provided features. Finally, in 3D, we
trace these feature edges as well to preserve surface features.

\subsection{Implementation Details}
Our pipeline is implemented using a mix of MATLAB~\cite{matlab} and C++ functions.
Notably, we rely on C++ finite element analysis code and some underlying C++
functions of the libigl library~\cite{libigl} called via the matlab gptoolbox code
base~\cite{gptoolbox}. Our frame fitting, texture parametrization, and mesh extraction
algorithms are implemented entirely in MATLAB\@. We use MATLAB's \texttt{fmincon} to solve
our frame fitting optimization and \texttt{quadprog} to solve for the texture paramaterization. We pledge to release all code and data pertaining to this submission as opensource.

\ud{\begin{figure}[H]
	\includegraphics[width=\columnwidth]{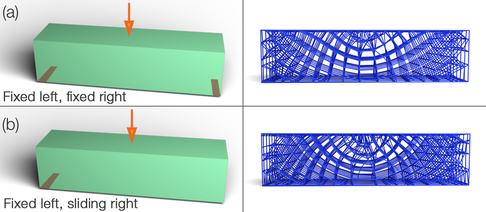}
	\caption{A point compression is applied at the center of the top face of the
	bar, with two different boundary conditions: fixed left and right edges on
	the bottom face (top), and fixed left edge with right edge allowed to slide
	horizontally (bottom). Notice the change in the slope of the curves with the 
	change in the boundary conditions.
	\label{fig:resultsSSFixed}}
\end{figure}
\section{Geometry Creation}
\ud{The final stage in our pipeline is to create geometry from our extracted truss networks. To do this we
specify a radius for each integer isocurve. We then replace each piecewise linear segment of the truss graph with a triangulated cylinder. The union of these cylinders, computed using libigl's~\cite{libigl} robust boolean operations, results in a full 3D geometry}  

We show virtual renders of the optimized trusses produced by our method in
Figures~\ref{fig:resultsMain} and ~\ref{fig:results_engg}. In all the figures in the paper, the input geometry 
is consistently shown in green, the forces applied with orange arrows, the
fixed points in maroon, and the locations of load application with a striped orange
pattern. For 3D results optimized purely under gravity, the ground contacts are
always set to be fixed, and are not explicitly shown in the figures.

Constructing volumetric Michell Trusses can be useful for many engineering applications.
\autoref{fig:results_engg} shows some applications of our method in the aerospace industry.
We also show furniture design applications (\autoref{fig:resultsGeometry}).

We also performed FEM analysis to visualize the stress fields in several of our examples (\autoref{fig:resultsSim}). 

\begin{figure}[htp]
	\includegraphics[width=\columnwidth]{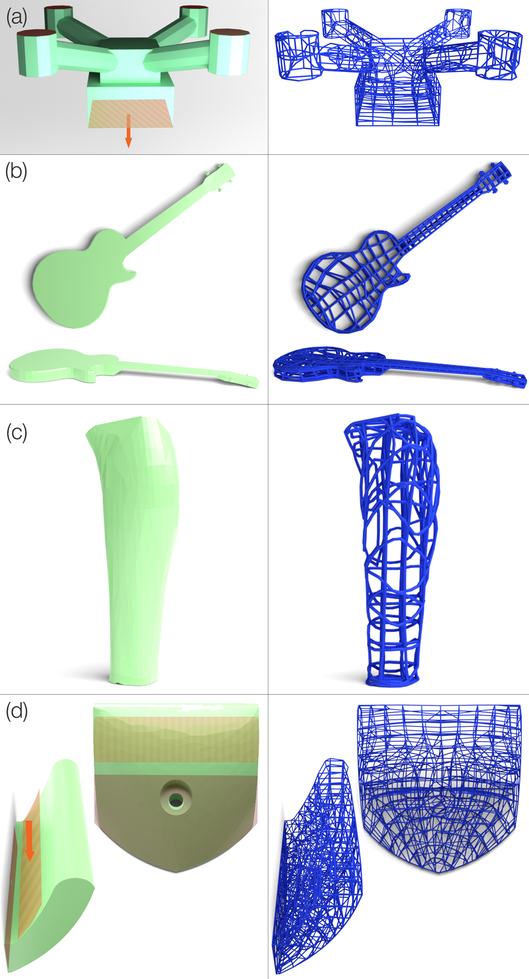}
	\caption{A selected set of trusses produced by our method. (Top to bottom)
	quadcopter fixed at the propeller mounts and optimized for carrying a parcel
	hanging from the bottom face, a guitar optimized for load applied by a guitar strap, a prosthetic leg and a rock climbing hold 
	(shown in two views for clarity). The left column shows inputs, and the 
	right column shows the corresponding rendered results.}
	\label{fig:resultsMain}
\end{figure}

\begin{figure}[htp]
	\includegraphics[width=\columnwidth]{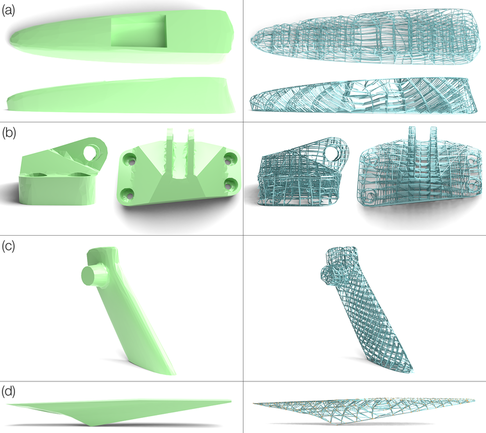}
	\caption{Aerospace applications: helicopter top pylon, airplane engine bracket, 
	helicopter rear pylon, and airplane engine/turbine pylon.
	\label{fig:results_engg}}
\end{figure}

\section{Expressivity of Michell Trusses}
In this section we explore the effect of geometry and boundary conditions on the output of our method. Figure~\ref{fig:resultsSSFixed} shows the results of our algorithm on two bars. The top bar has fixed Dirichlet boundary conditions at each end while the bottom allows free sliding of its right-hand side. Note the difference in the produced trusses.

Figure ~\ref{fig:resultsForce} shows the effect of changing the loading conditions. Two identical bars are subjected to twisting and compression respectively. The resulting truss structures adapt to provide maximum structural strength. 

\begin{figure}[H]
	\includegraphics[width=\columnwidth]{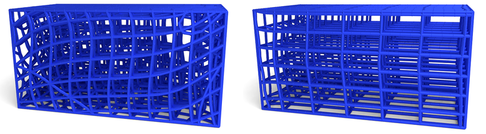}
	\caption{The difference caused by varying Neumann BCs, using
	bars under torsion (left) and compression (right). 
	\label{fig:resultsForce}}
\end{figure}

Figure~\ref{fig:resultsGeometry} shows two chairs with the same loading condition applied. The left chair is a more standard, swivel chair design while the right chair is produced using generative design software. Note the differences in the trusses due to the changes in geometry. 
 
\begin{figure}[H]
	\includegraphics[width=\columnwidth]{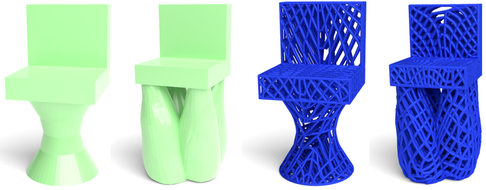}
	\caption{Two chairs under identical loading but with different geometries.
	\label{fig:resultsGeometry}}
\end{figure}

\begin{figure}[htb]
\centering
	\includegraphics[width=\columnwidth]{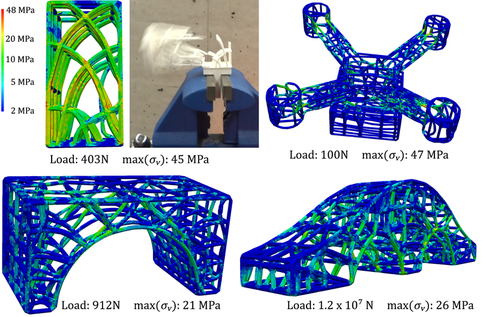}
	\caption{We used linear FEM to simulate our trusses. The material assumed for the visualizations is ABS-M30 plastic (yield strength 48 MPa), which we used for fabricating the bridge, quadcopter frame, and bending bar. The bending bar yielded at 403N. The simulation (a) shows excellent accuracy in predicting the stress concentration regions, which agree with experimentally observed fracture regions (b). The quadcopter frame is predicted to hold up 100N (10.2kg) of load successfully (c). The miniature (20cm wide) flat bridge simulation (d) predicts no fracture at 912N (93kg) load. A real-world scale arch-bridge made of 2.5cm thick elements (c) is predicted to hold $1.2\times10^7$ N (10 firetrucks).
	\label{fig:resultsSim}}
\end{figure}

\begin{figure}[tbh]
\centering
\includegraphics[width=\columnwidth]{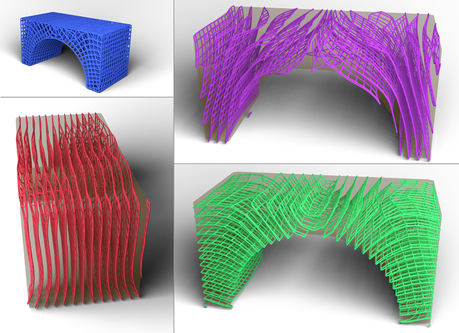}
\caption{Visualization of the trusses extracted by our method by tracing the 
integer isosurfaces of $\tphi$. The full truss for the bridge problem
(blue), and truss elements traced on integer isosurfaces of each $\tphi_i$
in red, green, and violet.}
\label{fig:bridgeIsosurfaceViz}
\end{figure}


\section{Performance}
\label{sec:performance}

Table~\ref{tbl:performance} reports the performance of our algorithm for all the
tested 3D problems---problem size, total running time, and the time taken by
individual steps of the algorithm. The results indicate running time on a 2017 Macbook Pro with Intel Core i7 Processor and 16 GB of RAM. Note that most of our code is written in MATLAB and is not optimized for
speed. Based on previous studies, we expect an order of magnitude or more
speedup simply by implementing the algorithm in C++~\cite{kjolstad:2016}.
\begin{table*}[htb]
 	\centering
	\rowcolors{2}{white}{gray!25}
  	\begin{tabular}{lccccccccc}
    	\midrule
    	\textbf{Example} &     \textbf{$\#$ Vertices} & \textbf{$\#$ Tets} &    \textbf{Total (s)} & \textbf{Sim (s)} &    \textbf{Frames (s)} & \textbf{Tex (s)} & \textbf{Extract (s)}\\
	\midrule
	Bar (bending)&3457 & 16704 & 618.0 & 0.5 & 20.8 & 1.2 & 595.4\\
	Bar (simply supported)&6913 & 34128 & 583.1 & 1.3 & 69.1 & 3.7 & 509.0\\
	Climbing Hold&13753 & 68773 & 2189.7 & 3.7 & 209.8 & 14.8 & 1961.5\\
	Guitar&10171 & 46840 & 689.9 & 1.3 & 54.1 & 7.2 & 627.2\\
	Bar (torsion)&3457 & 16704 & 2239.2 & 0.4 & 17.3 & 1.0 & 2220.4\\
	Bar (compression)&3457 & 16704 & 1411.4 & 0.5 & 15.5 & 1.0 & 1394.4\\
	Chair (swivel)&7372 & 35021 & 4587.1 & 1.2 & 102.7 & 4.0 & 4479.3\\
	Chair (generative)&9801 & 46187 & 6278.7 & 1.7 & 99.5 & 6.7 & 6170.8\\
	Bar (fixed)&6913 & 34128 & 511.2 & 1.2 & 51.0 & 3.6 & 455.4\\
	Pylon (helicopter, top)&8268 & 35449 & 831.9 & 1.0 & 44.4 & 4.3 & 782.3\\
	Pylon (helicopter, rear)&4038 & 17358 & 609.4 & 0.3 & 17.8 & 1.2 & 590.0\\
	Prosthetic&9097 & 44478 & 1380.3 & 1.5 & 77.2 & 5.9 & 1295.6\\
	Bridge&4326 & 18808 & 648.5 & 0.5 & 89.8 & 1.5 & 556.6\\
	Pylon (turbine)&2520 & 8932 & 225.9 & 0.2 & 8.0 & 0.4 & 217.3\\
	Antenna arm&3264 & 13459 & 1345.6 & 0.3 & 13.0 & 0.7 & 1331.6\\
	Jet engine bracket&28904 & 135649 & 3996.4 & 4.0 & 1501.2 & 52.1 & 2439.1\\
	Bunny&19866 & 87443 & 1471.2 & 3.7 & 175.3 & 23.6 & 1268.6\\
	Quadcopter&7121 & 32826 & 671.2 & 1.0 & 45.9 & 3.7 & 620.7\\
	Bridge (arch)&6457 & 31164 & 621.2 & 0.9 & 73.7 & 3.1 & 543.5\\
\midrule
    \end{tabular}
    \caption{Performance results for all the 3D testcases. All reported runtimes have been rounded off to the nearest second.}
    \label{tbl:performance}
\end{table*}


}

\section{Conclusion}\label{sec:conclusion}
By adopting a parametrization-based approach we have crafted the first algorithm
for the design of volumetric Michell Trusses and shown that the algorithm can produce
complex, aesthetically pleasing output that is also strong.
We believe
our method serves as an important companion to traditional approaches while also
providing engineers, architects, and designers with an exciting new algorithmic tool.

\paragraph{Limitations and Future Work.} 
\ud{The most significant limitation of our approach is that it does not incorporate fabrication constraints. 
Incorporating such constraints into design optimization is an ongoing area of research. We are motivated by recent works to investigate this further~\cite{Allaire:2016:Con,martinez:hal:poly}.}

\begin{figure}[h]
	\includegraphics[width=\columnwidth]{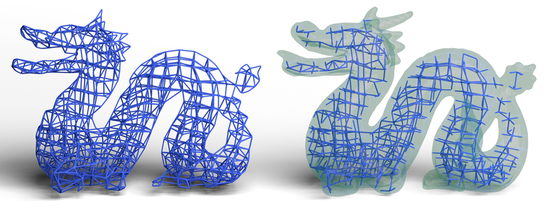}
	\caption{The trusses produced by our method may fail to cover thin features
	like the dragon's horns (left). However, it is still useful for optimizing
	the interior of an object meant to be fabricated with a solid surface (right).}
	\label{fig:limitationDragon}
\end{figure}

Our method produces trusses whose
members are almost equally spaced.
This can be problematic when the input geometry contains
very thin regions, such as the dragon shown in
Figure~\ref{fig:limitationDragon}, where we completely miss the dragon's horns.
Fortunately, if the aim is to build objects with a solid shell as the surface,
we can still optimize the internal structure.

In general, efficient and fine-grained user control of topology optimization
remains an interesting direction for future research.

\setlength{\intextsep}{0pt}
\setlength{\columnsep}{0pt}
\begin{wrapfigure}[8]{r}{0.3\columnwidth}
  \begin{center}
      \includegraphics[width=0.3\columnwidth]{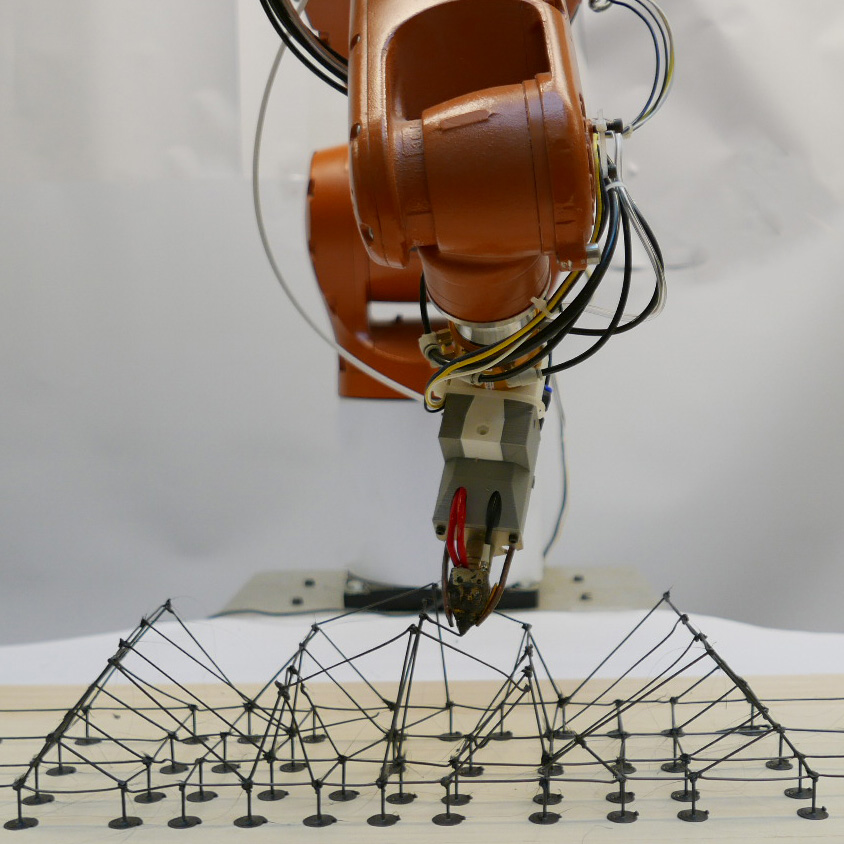}
  \end{center}
\end{wrapfigure}
\setlength{\intextsep}{3pt}
\setlength{\columnsep}{10pt}
Lastly, parameterized trusses can be used to fabricate structures using
new, exotic, fabrication methods such as extruders mounted on robot arms with many
degrees of freedom~\cite{Tam:2015:Stress, MX3D:2017:Metal}. Such robot arms can build large, self-supported
structures in a fraction of time required by current generation 3D printers, and 
therefore, hold promise to be the additive manufacturing technique of the future.


\bibliographystyle{ACM-Reference-Format}
\bibliography{artopt}

\end{document}